\documentclass[RNAAS]{aastex631}

\begin{document}

\title{On the Measurement of Vorticity in Astrophysical Fluids}

\correspondingauthor{Steven Spangler}
\email{steven-spangler@uiowa.edu}

\author[0000-0002-4909-9684]{Steven R. Spangler}
\affiliation{Department of Physics and Astronomy, University of Iowa}

\begin{abstract}
Vorticity is central to the nature of, and dynamical processes in turbulence, including turbulence in astrophysical fluids.  The results of \cite{Raymond20a,Raymond20b} on vorticity in the post-shock fluid of the Cygnus Loop supernova remnant are therefore of great interest.  We consider the degree to which spectroscopic measurements of an optically-thin line, the most common type of astronomical velocimetry, can yield unambiguous measurements of the vorticity in a fluid.  We consider an ideal case of observations in the plane of a flow which may or may not contain vorticity.  In one case, the flow possesses vorticity in a direction perpendicular to the plane of observations.  In the other case, the flow is irrotational (zero vorticity) by construction.  The observationally-deduced vorticity (referred to as the {\em pseudovorticity}) is inferred from spatial differences in the line-of-sight component of velocity, and assumptions of symmetry. My principal result is that in the case of the vortical flow, the pseudovorticity is a reasonable match for the true vorticity.  However, and importantly, the pseudovorticity  in the case of the irrotational flow field is also nonzero, and comparable in magnitude to that for a vortical flow.  The conclusion of this paper is that while astronomical spectroscopic observations may yield a good estimate of the vorticity in a remote fluid, the robustness of such an inference cannot be insured.
\end{abstract}
\keywords{interstellar medium --- turbulence --- interstellar plasma}

\section{1. Introduction} 
\begin{center}
``Zeus no longer exists, a {\em Vortex} now rules in his place ''  \ldots  Aristophanes, ``The Clouds''  
\end{center}
Turbulence is believed to play an important, and perhaps dominant role in many astrophysical processes. A full litany of such processes would take a great deal of space, and has been given in many other publications.  However, to cite just one example, turbulence is believed to play an important role in modulating star formation \citep{Guszejnov20}.  It may be important to distinguish between turbulence in plasmas, for which the Lorentz force is a dominant term, and turbulence in neutral fluids (hydrodynamic turbulence), in which that term is absent.  However, the most common theoretical description of plasmas used in astrophysics, single fluid magnetohydrodynamics (MHD), contains all the terms present in the Navier-Stokes equations.  One therefore expects close similarities between hydrodynamic and magnetohydrodynamic turbulence.

Vorticity is crucial to the nature and dynamics of turbulence, a fact apparently recognized by the Pre-Socratic philosophers of Ancient Greece, and conveyed in the above excerpt from {\em The Clouds} of Aristophanes, written around 423 BCE \footnote{This phrase was so important to the theme of ``The Clouds'', that Aristophanes included it three times in the comedy.}. In terms of a modern mathematical description of fluid mechanics, the vorticity $\vec{\omega}$ is related to the fluid velocity field $\vec{v}(\vec{r},t)$ as
\begin{equation}
  \vec{\omega} = \nabla \times \vec{v}
\end{equation}
where $\vec{r}$ is the vector position and $t$ is the time.

Descriptions of hydrodynamic turbulence generally envision the turbulent field as a superposition of vorticies of differing sizes and intensities \citep[e.g][]{Bradshaw78,Vincent91}. Furthermore, the amplification of vorticity in the strain field 
\begin{equation}
S_{ij} \equiv \frac{1}{2} \left( \frac{\partial v_i}{\partial x_j} + \frac{\partial v_j}{\partial x_i} \right)
\end{equation}
 of the surrounding flow is considered the dominant mechanism for the famous turbulent cascade from large scales to small scales.  As clearly explained in \citet{Bradshaw78} a vortex on one spatial scale is stretched and amplified by the flow field of vorticies on  larger scales.

It should be noted that this view of ``vortex stretching'' as the sole mechanism causing the turbulent cascade may need revision.  Recent work \citep{Johnson21,Carbone20} has indicated that ``strain amplification'' can compete with, and perhaps dominate, vortex stretching as a cascade process.  Strain amplification can simplistically be described as enhancement of gradients due to convergent flows, and is familiar from one dimensional nonlinear wave equations such as Burger's Equation and the Derivative Nonlinear Schroedinger Equation \citep[e.g.][]{Spangler85}. However, it may be the case that vorticity dynamics are responsible for highly intermittent events in turbulence \citep[ see ][]{Johnson16}.

A variety of astronomical measurements have shown that turbulence exists in media such as the Solar Corona, all phases of the Galactic Interstellar Medium, and perhaps extragalactic radio sources.  However, the diagnostics available to astronomers, such as radio wave scattering from density fluctuations \citep{Rickett90} or turbulently-broadened spectral lines \citep{Redfield04} are not suited for insight into the dynamics of the turbulence. The case of the Solar Wind is in much better shape, in that impressive, in-situ measurements of flow velocity, magnetic field, and density fluctuations can be made over a wide range of spatial scales \citep{Tu95,Bruno05}.  Even here, however, inferring quantities such as the vorticity and determining the turbulent dynamics are very difficult \citep[see, however ][]{Andres21}.

Given the above remarks, it would seem virtually impossible to measure the vorticity, rate-of-strain tensor, or other such quantity in remotely-sensed astrophysical media.  For this reason, the recent results of \cite{Raymond20a,Raymond20b} were a cause of great enthusiasm for astronomical ``cumaphilonts'' \footnote{A Hellenistically-motivated term meaning ``lovers of turbulence'' that I originally introduced in \cite{Spangler01}, without any noticeable success.}. Utilizing spectroscopic and proper motion measurements of post-shock flow in the western portion of the Cygnus Loop, \cite{Raymond20a,Raymond20b} report detection and measurement of vorticity in that flow. \cite{Raymond20a,Raymond20b} were clear and candid in the assumptions that were used in their vorticity detection.  In particular, assumptions were made about the symmetry of the flow that could not be subject to observational verification or falsification.  

The results of \cite{Raymond20a,Raymond20b} motivated the present investigation. I investigate the {\em spectroscopic} inference of vorticity for a set of idealized, model flows.  I restrict attention to spectroscopic measurements of flow Doppler shifts, because that is the most common method of astronomical anemometry. The results presented in \cite{Raymond20a} used only spectroscopically-determined Doppler shifts, and thus fall in this category. Proper motion measurements as utilized by \cite{Raymond20b} are an exception to this statement, as are some types of radio scintillation measurements that measure the velocity at which a scintillation pattern moves across a radio telescope array \citep[ e.g.][]{Coles91}.  The goals of the present investigation are to determine (1) if spatially-varying Doppler shifts measured spectroscopically in the case of a true vortical flow allow an accurate estimate of the vorticity to be retrieved, and (2) if flows with $\vec{\omega} = 0$ by construction would yield a spectroscopic signature which would be misinterpreted as vorticity.

The organization of this paper is as follows.  In Section 2, I make a cursory inspection of laboratory measurements of vorticity in hydrodynamic turbulence.  The point of including this section, other than its intrinsic interest, is to acknowledge that even in controlled laboratory settings with internal and external probes, measurement of vorticity is difficult.  Section 3 considers the flow associated with a vortex being amplified by a strain along the vortex axis, referred to as the {\em Batchelor Flow} \citep{Batchelor67}. I then discuss the conversion between a 3 dimensional flow field  $\vec{v}(\vec{r})$ and the line-of-sight component (radial velocity) available in astronomical Doppler measurements. I describe how spectroscopic observations of spatially-varying Doppler shifts in an astronomical spectroscopic mapping experiment might be used to estimate the vorticity. Since this is an observational proxy for an unmeasureable physical quantity, I call this quantity the {\em pseudovorticity}. Section 4 presents similar analysis of a flow field chosen by construction to have $\vec{\omega} = 0$ everywhere. My conclusion is that a similar pseudovorticity would be inferred for both flow fields, although the true vorticity is entirely different. Section 5 points out that these conclusions could have been anticipated from clear insights on the nature of Galactic hydrogen flows presented by W.B. Burton in a set of articles approximately 50 years ago. Section 6 summarizes and concludes.  
\section{2. The Measurement of Vorticity in Laboratory Turbulence}
An interesting aspect of the study of turbulence is its importance, not only to fundamental physics, but also to research in the engineering community.  Many of the most insightful books and journal articles have been written by engineers.  Given that this phenomenon has attracted the scrutiny of flinty-eyed engineers, one would expect more solid evidence on fundamental questions than typically can be provided by physicists, let alone astronomers. One would expect the existence and dynamical importance of vorticity in hydrodynamic turbulence to be well-established experimentally.  Curiously, this seems not to be the case.

Measurement of turbulent fluid motions, including vorticity, are made utilizing a number of experimental techniques, such as hot-wire anemometry \citep{Hubbard57,Comte76,Wallace95}, optical tracking of a passive tracer \citep{Wallace09,Wallace10}, and particle image velocimetry \citep{Wallace09,Wallace10}.  Perhaps the most standard method is that of hot-wire-anemometry, which consists of measuring the cooling rate of a heated wire in a moving fluid.  An empirical, though theoretically-based expression for the response of a hot-wire anemometer is \citep{Hubbard57}
\begin{eqnarray}
  \frac{I^2 R \alpha}{a} = C_1 + C_2 \sqrt{U} \\
  a \equiv \frac{R-R_0}{R_0}
\end{eqnarray}
where $I$ is the current flowing through the circuit, $U$ is the component of the flow velocity perpendicular to the wire, $R$ is the actual resistance of the wire, $R_0$ is a reference value for the resistance of the wire (defined as the resistance when the resistor has the temperature of the fluid), and $\alpha$ is the thermal conductivity of the fluid.  The coefficients $C_1$ and $C_2$ are constants characteristic of the wire, such as platinum, that is employed in the device. Measurement of $U$ results from measuring the current $I$ necessary to maintain the resistance of the wire at $R$.  

A hot wire anemometer is not a vector instrument. The cooling rate, and thus the resistance $R$ depend on the angle between the wire and the flow direction.  An array of wires oriented at different angles can deduce the vector flow velocity at a point.  An array of at least 9 hot-wire anemometers, spatially-separated and oriented at specifically chosen directions, can, in principle, measure the vorticity at a point in the fluid.  Such a hot-wire anemometer array is called a ``Kovasznay probe'' after its inventor, and is sometimes referred to as a ``vorticity meter''.

Nonetheless, technical problems limit the capability of measuring vorticity with hot-wire anemometer arrays.  The size of the individual elements (wires) is problematic.  A wire which is too big will average over important spatial scales, while a wire which is too small will have significant noise error. In addition, the structures which hold the wires in place within the flow will modify the flow field from what it would be in the absence of those structures.  Discussions of these issues may be found in \cite{Comte76}, \cite{Wallace95,Wallace09,Wallace10}.

Studies of turbulence with hot-wire anemometer arrays have yielded some results which are broadly consistent with the presence of vorticies.  \cite{Kuo72} found that the regions of fine-scale fluid fluctuations tended to be in geometric structures which were ``rods'', subsequently interpreted as ``tube-shaped regions of concentrated high vorticity'' \citep{Johnson16}.  

A recent, and extremely clever technique was described by \cite{Wu19}.  Tiny, micron sized spheres are fabricated from  a material with nearly the same index of refraction as water, and possessing embedded mirrors. A number of these spheres are then seeded into a turbulent flow of water.  The vorticity equals twice the angular velocity at a point.  Scattering of a light ray from a laser, and off the spheres then can measure the vector angular velocity, and thus the vorticity, at a point in the fluid. \cite{Wu19} verify the technique by showing that it yields the theoretically-expected distribution of vorticity in a Couette flow, but to date the technique does not seem to have been used to obtain results in large-inertial subrange turbulence.

The incomplete state of knowledge of vorticity in laboratory turbulence is conveyed by a statement in the introduction of \cite{Wu19}: ``The vorticity field underlies the dynamics of turbulent flows.  Measuring vorticity experimentally, however, is notoriously difficult''.

In concluding this brief and incomplete sampling of the literature on laboratory measurements of turbulence, I would emphasize the reason for including this material at all.  If, as appears to be the case, it is difficult to clearly measure and prove the presence of vorticity in laboratory flows, where there is control over the flow conditions and geometry, one can reasonably expect a much more difficult problem in the case of astronomical fluids, for which incomplete, remote sensing measurements are the only ones available.

\section{3. The Batchelor Flow: A Model Flow with High Vorticity}
In this section, I consider a flow field known to possess vorticity as well as vortex stretching, so the vorticity increases with time.

Starting with the Navier-Stokes equations, the following equation for the vorticity can be derived \citep{Bradshaw78}
\begin{equation}
  \frac{\partial \vec{\omega}}{\partial t} + (\vec{v} \cdot \nabla)\vec{\omega} = (\vec{\omega} \cdot \nabla)\vec{v} + \mbox{ viscous terms }
\end{equation}
The terms on the left hand side constitute the familiar convective derivative, while the term on the right hand side describes an ``amplification'' of vorticity due to the strain field.  The omitted viscous terms \citep{Bradshaw78} describe spatial diffusion of vorticity due to viscosity. This equation shows the basic physics content of vortex amplification by stretching; when there is a positive derivative of the component of the flow velocity in the direction of the vorticity, the vorticity is amplified. A clear discussion is in \cite{Bradshaw78}.

Since a vortex consists of ``swirling'' motion, for the following discussion I utilize cylindrical coordinates ($R$, $\phi$, $z$), and take the direction of the vorticity to be in the z direction.  
\subsection{3.1 The Batchelor Flow}
\cite{Batchelor67} considered a flow field which would lead to vortex stretching, and a vorticity field given by
\begin{equation}
  \vec{\omega} = 0 \hat{e}_R +  0 \hat{e}_{\phi} + \omega(R,t) \hat{e}_z
\end{equation}
The conditions for the vorticity to be of this form are
\begin{equation}
  \frac{\partial v_{\phi}}{\partial z } =  \frac{\partial v_R}{\partial z} =  \frac{\partial v_z}{\partial R} = 0
\end{equation}
The background flow field chosen by \cite{Batchelor67}, which satisfies the above relations, is
\begin{eqnarray}
  v_z = \alpha z \\
  v_R = -\left( \frac{\alpha}{2} \right) R 
\end{eqnarray}
where $\alpha$ is a constant with dimensions of $t^{-1}$.  The azimuthal component of the velocity, $v_{\phi}$ is contained in the vortex being amplified.  Substitution of this background flow into Equation (5) yields Equation 5.2.9 of \cite{Batchelor67}, an expression for the scalar vorticity $\omega$
\begin{equation}
   \frac{\partial \omega}{\partial t} = \left( \frac{\alpha}{2} \right) \frac{1}{R} \frac{\partial}{\partial R} \left( R^2 \omega \right)
 \end{equation}
 Equation (10) omits a term describing viscous diffusion of vorticity. 
  \subsection{3.2 A Specific Solution}
  Given Equation (10), I am looking for a solution that will provide a specific form of $\omega(R,t)$, that can then be converted to a form for $v_{\phi}(R,t)$.  I choose a separable form,
  \begin{eqnarray}
    \omega(R,t) = A(t) w(x) \\
    x \equiv \frac{R}{a(t)}
  \end{eqnarray}
  where $a(t)$ is a time-dependent scale factor for the vortical flow.  
  This form has the Kelvin Circulation Theorem, with conservation of the circulation, in mind from the start.  
  Substituting this trial function $ \omega(R,t)$ into Equation (10), the following form emerges
  \begin{equation}
    \dot{A}w - Aw^{'} R \left( \frac{\dot{a}}{a^2} \right) = \left( \frac{\alpha}{2} \right) A \left[ 2w + \left( \frac{R}{a} \right) w^{'} \right]
  \end{equation}
  where $\dot{A} = \frac{dA}{dt}$ and $w^{'} =  \frac{dw}{dx}$.
  Dividing through by $Aw$ gives
  \begin{equation}
    \frac{\dot{A}}{A} - \left( \frac{w^{'}}{w} \right) R \left[ \frac{\dot{a}}{a^2} + \left( \frac{\alpha}{2} \right) \frac{1}{a} \right] = \alpha
  \end{equation}

  Any solution to this equation for the functions $A(t)$ and $w(x)$ is a candidate physical solution, if it behaves in a physical way and can match appropriate initial conditions.

  One obvious solution is to set the term in square brackets to zero, i.e.
  \begin{equation}
  \left[ \frac{\dot{a}}{a^2} + \left( \frac{\alpha}{2} \right) \frac{1}{a} \right] = 0  
\end{equation}
giving the solution for $a(t)$
\begin{equation}
  a(t) = a_0e^{-\alpha t/2}
\end{equation}
corresponding to a vortex which contracts with time.  If this solution is used, the resulting expression for $A(t)$ is
\begin{equation}
  A(t) = A_0 e^{\alpha t}
\end{equation}
corresponding to a strengthening of the vortex as it contracts. Equations (16) and (17) suggest the introduction of a dimensionless time $\tau \equiv \alpha t$.  
Given these restrictions on the strength and scale of the vortex as it is amplified, any form for the function $w(x)$, which describes the spatial distribution of the vorticity, is acceptable.  A physically plausible choice is one which has a maximum on the axis of the vortex, and decreases with increasing radial distance from the axis. The obvious choice is a Gaussian,
\begin{equation}
  w(x) = e^{-x^2}
\end{equation}
with dimensional factors collected into the normalization constant $A_0$.
Equations (16), (17), and (18) can be used to constitute an expression for the time and space-dependent vorticity,
\begin{equation}
  \omega(R,t) = \omega_0 e^{\tau} \exp \left[ -\left( \frac{R}{a_0} \right)^2  e^{\tau}  \right]
\end{equation}
where $\omega_0$ is the initial vorticity on axis, and $a_0$ is the initial radius of the vortex (defined in Equation (16)). 
This solution will be used for the vortical flow in the remainder of this investigation.

Given the above expressions for the vorticity $\vec{\omega}(R,t)$ we want to obtain an expression for the flow field in the vortex, $\vec{v}(R,t)$.  From the definition of the vorticity and Equations (6) and (7) we have
\begin{equation}
  \vec{\omega} = \omega(R,t) \hat{e}_z = \frac{1}{R} \frac{\partial}{\partial R} \left( R v_{\phi} \right) \hat{e}_z
\end{equation}
The vortical flow, $v_{\phi} (R,t)$ is azimuthally symmetric and invariant in z.

If Equation (11) is substituted into Equation (20), and integrated from $0$ to $R$, the following expression for $v_{\phi}$ results
\begin{equation}
  v_{\phi}(R,t) = \left[ \frac{\omega_0 a_0^2}{2} \right] \left( \frac{1}{R} \right) \left[ 1 - e^{-R^2/a^2}  \right]
\end{equation}
where the constant $A_0$ in Equation (17) is identified with the initial vorticity $\omega_0$.  The Kelvin Circulation Theorem is expressed by the fact that $\omega_0 a_0^2$  is a constant.

The solution given by Equation (21) has the property that for $R \gg a$, $v_{\phi} \propto R^{-1}$, which is correspondent to the Biot-Savart Law in electricity and magnetism.  In this part of space, the vorticity is approximately zero, supporting the physical relevance of the solution adopted in Equation (15).

In comparing the flow field Equation (21) to synthetic observations, it will be helpful to have fixed, rather than time variable, dimensionless coordinates.  I therefore pick a new dimensionless variable y,
\begin{equation}
  y \equiv \frac{R}{a_0}
\end{equation}
The flow field is then given by
\begin{equation}
   v_{\phi}(y,t) = \left[ \frac{\omega_0 a_0}{2} \right] \left( \frac{1}{y} \right) \left[ 1 - \exp{ (-y^2 e^{\tau})}  \right] 
 \end{equation}
 where $\tau \equiv \alpha t$ is the dimensionless ``age'' of the stretched vortex.  I also define a new constant $V_0 \equiv \frac{\omega_0 a_0}{2}$ as the velocity scale associated with the vortex. The expression in the last term on the right hand side of Equation (23) shows the profile of vorticity in the stretched and contracting vortex.  Assumed expressions for $w(x)$ different from that adopted in Equation (18) would presumably differ in the expression in this term.

 The velocity profile corresponding to Equation (23) is shown in Figure 1 for three values of the dimensionless time $\tau$.
 \begin{figure}[h!]
\begin{center}
\includegraphics[scale=0.85,angle=0]{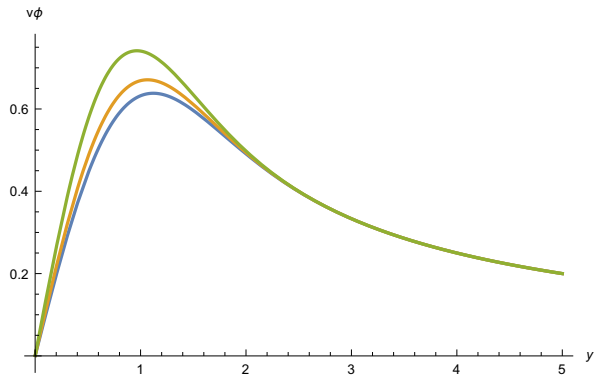}
\caption{The velocity profile $v_{\phi}$ as a function of dimensionless distance $y$ from the vortex axis for three different dimensionless times $\tau = \alpha t = 0, 0.1, 0.3$ (blue, yellow, green, respectively).  The azimuthal velocity reaches a peak for $y \sim 1$, then asymptotes to a irrotational, $R^{-1}$ form for sufficiently large values of y.  This plot shows that with increasing time $\tau$, the vortex contracts and the azimuthal velocity increases.}
\end{center}
\end{figure}

\subsection{3.3 The Optically Thin Spectrum from a Batchelor Flow}
In most cases, astronomical anemometry consists of the measurement of a Doppler-shifted spectral line at a given position on the sky.  The spectrum is formed, in the optically-thin case, by contributions of emitting elements all along the line of sight. Velocity measurements therefore consist of a weighted average velocity along a line segment from the observer to infinity in one direction.  This is a dramatic contrast with laboratory measurements which seek to measure the fluid velocity $\vec{v}(\vec{r})$ at a given point $\vec{r}$.  Attempts to determine spatial functions of the velocity, such as the vorticity, therefore require consideration of the velocity profile along the line of sight, if such is knowable.  In this section, I consider the ``mapping'' between the velocity field and the observed spectrum for the case of the Batchelor Flow.  In Section 4, I do the same for model flow fields chosen to be irrotational.

A number of drastic approximations are made to simplify the analysis.  These could be relaxed in future investigations, but my reason for making them is that with these approximations, the vortical signature should be at a maximum.  Relaxation of these restrictions, and consideration of a flow field possessing a number of vortices (the obvious realistic case), would presumably blur the signature of vorticity, although this assertion should be verified.

The assumptions and restrictions employed are as follows.
\begin{enumerate}
\item I assume that the line radiation emitted by the fluid in the vortex is optically-thin.  This permits the intensity of the radiation to be expressed as a line integral along the line-of-sight.  In astrophysical applications, this is often the case, but not always.
\item I assume that line-of-sight lies entirely in a plane with a constant value of the cylindrical coordinate $z$.  This means that the axis of the vortex is perpendicular to the line-of-sight.  In this plane, we introduce a Cartesian coordinate system $(s,t)$ (t not to be confused with the time), where $t$ is a coordinate along the line-of-sight, and $s$ is a coordinate in the plane of the sky, and perpendicular to the axis of the vortex. This coordinate system is illustrated in Figure 2. This assumption clearly simplifies the analysis, since in general, there will be two angles describing the orientation of the vortex axis with respect to the plane of the sky.  The restriction chosen should maximize the ``signal'' due to the vortex.
\item I also assume that the velocity field measured in the spectrum is exclusively due to the velocity field in the vortex (Equation (23)), i.e. the vortex dominates over the background fluid flow ( Equation (8) and (9)). The $z$ component of the background flow will not contribute because of the restriction \# 2 above, but $v_R$ will, in general, contribute to the spectrum. My assumption will be valid if the background, ``stretching'' flow is slow compared with the flow in the vortex.  This assumption will always be true if the vortex has grown sufficiently.
\item The finite extent of the emitting region, taken to be larger than the size of the vortex, is parameterized by a length scale $L$ in the $t$ direction.  The density of the emitting atoms or molecules is taken to be uniform for $-L/2 \leq t \leq L/2$ and zero outside.  The constant density is compatible with incompressible flow.
  \item Given the coordinate system defined in item \# 2 and illustrated in Figure 2, the Doppler shift of a fluid element depends only on the velocity component $v_t$, will vary with the $t$ coordinate, and be symmetric about $t=0$.  The spectrum will be determined by the value of $s$, and change with the value of $s$. 
  \end{enumerate}
 \begin{figure}[h!]
\begin{center}
\includegraphics[scale=0.85,angle=0]{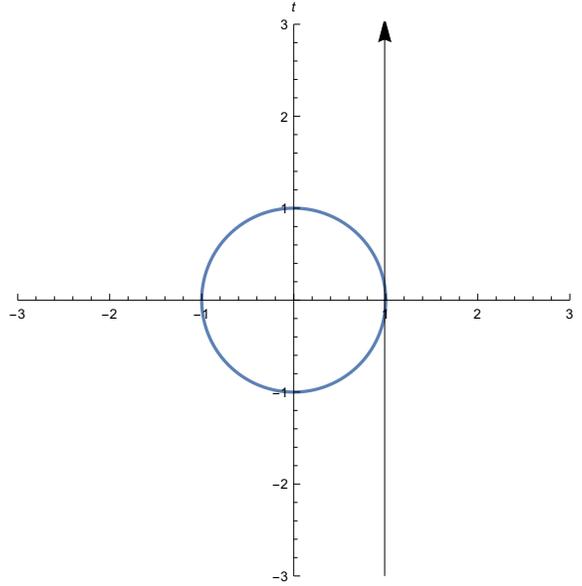}
\caption{Diagram showing line-of-sight (LOS) relative to the coordinate system used to describe the vortex. The s and t coordinates are (dimensionless) Cartesian coordinates in the $z=0$ plane of the vortex. The vertical arrow represents the LOS, which has all values of $t$ and a constant value of $s$.  The blue circle represents the cylindrical coordinate system used to derive the expression for the flow in the Batchelor Vortex, Equation (23).  This circle may be considered the locus of points with constant vorticity.}
\end{center}
\end{figure}
Given these assumptions and simplifications, the Doppler shift of a given fluid element is given by
\begin{equation}
  v_t(t(R,\phi)) = v_{\phi}(R) \cos \phi
  \end{equation}
where $v_{\phi}(R)$ is given by Equation (21), and the cylindrical coordinate $\phi = 0$ along the positive $s$ axis.  The spectrum at a given position $s$ contains a range of velocities from $v_t(\mbox{max}) \equiv v_{\phi}(R=s)$ to $v_t(\mbox{min}) \equiv v_{\phi}(R=R_{\mbox{max}}) \cos \phi_{\mbox{max}}$ 
where $R_{\mbox{max}}$ and $\phi_{\mbox{max}}$ are the values of the cylindrical coordinates at the point where the line-of-sight enters or exits the emitting region.

Using the definition of these parameters, one can convert from $v_t(t(R,\phi)) \rightarrow v_t(s,t)$.  I also introduce dimensionless versions of the Cartesian coordinates, choosing the initial vortex radius $a_0$ as the dimensional scale,
\begin{equation}
  s = a_0 s^{'}, t= a_0 t^{'}, L= a_0 L^{'}
\end{equation}
Substituting these expressions into Equation (23), and from now on noting the dimensionless coordinates by $(s,t)$ (i.e. dropping the primes), I have
\begin{equation}
  v_t(s,t) = V_0 \left[ \frac{s}{s^2 + t^2} \right] \left(1 - \exp[-(s^2 + t^2)e^\tau]  \right)
\end{equation}
for $-L^{'}/2 \leq t \leq L^{'}/2$, and again, $V_0 \equiv \frac{\omega_0 a_0}{2}$.  So,
\begin{equation}
  v_t = v_t(V_0, L^{'}, \tau, s, t)
\end{equation}
A plot of $ v_t$ for the Batchelor Flow (Eq. 25) is shown in Figure 3.  Curves are shown for three values of the dimensionless ``transverse'' coordinate $s$.
 \begin{figure}[h!]
\begin{center}
\includegraphics[scale=0.85,angle=0]{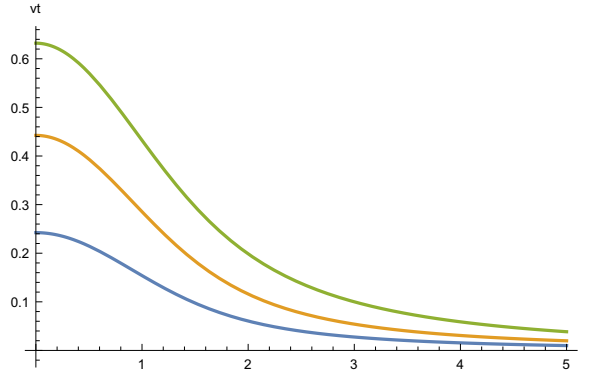}
\caption{The line-of-sight component of the fluid velocity ($ v_t$) in the vortex as a function of $t$, the coordinate along the line of sight.  For all three curves, the normalization constant of the velocity $V_0 = 1.0$, the dimensionaless time $\tau=0$, and the dimensionless thickness of the emitting region $L = 10$. The three curves give $v_t$ for spatial offset values (distance from the axis of the vortex, dimensionless) of $s=0.25$ (blue), $s=0.50$ (yellow), and $s=1.00$ (green).  In all cases, it is seen that the line-of-sight component of the velocity maximizes at $t=0$, and diminishes to a low value near the boundary of the emitting region. }
\end{center}
\end{figure}
It may be seen that for all three values of $s$, the maximum value of $v_t$ occurs at $t=0$.  This figure also shows that for each velocity within the spectrum, there are two points along the line of sight $\pm t$ with that value.  For sufficiently large values of $t \rightarrow L^{'}/2$, the velocity asymptotes to zero. Figure 3 also shows that the LOS component of the velocity, $v_t$, ranges from a maximum value of $v_t(\mbox{max}) = v_t(s,0)$ to a minimum value $v_t(\mbox{min}) = v_t(s,\pm L/2)$.  
\subsection{3.4 Relating the Velocity Field to the Emission Spectrum}
With a radio or optical telescope, one measures the intensity (or flux density if the source is unresolved) of emission as a function of frequency.  It is conventional in astronomy to change the frequency variable from the fundamental wavefrequency $f$ to the associated Doppler shift velocity $V$.

In astronomical sources, the LOS component of the fluid velocity will change with position along the line of sight, leading to a broadened spectrum. In section 3.3 above, I gave an expression (Equation 26) for the LOS component of the vortex velocity of a Batchelor Flow as a function of the coordinates $(s,t)$.  In the present section, I use this expression to obtain an equation for the (potentially observable) spectrum of a Batchelor vortex.  In the optically-thin case, the incremental change in the intensity as a function of Doppler velocity $V$ with LOS position $t$ is\footnote{For clarity of presentation, I use the variable $V$ to indicate a Doppler-shifted velocity as would be measured at a telescope, to distinguish it from $v_t$, which is the LOS component of the fluid velocity at a given position in the fluid.}
\begin{equation}
  dI(V) = j(V,t) a_0 dt
\end{equation}
where again, $t$ is a dimensionless coordinate (Figure 2), and $j(V,t)$ is the emission coefficient.

It is reasonable to assume that the intrinsic line width of the radiation is much smaller than the spectral broadening due to velocity variations along the LOS.  If the gas properties are otherwise uniform in the whole region for $-L/2 \leq t \leq L/2$, then changes in the emission coefficient are due to changes in the Doppler velocity $V$ at which the emission occurs, so
\begin{equation}
  j(V,t) = \frac{J_0}{a_0} \delta(V - v_t(t))
\end{equation}
where the normalization coefficient $J_0$ (a measure of the absolute intensity of the radiation) accommodates the dimensional factors needed in the use of Dirac delta functions. Thermal broadening, of course, will generally be larger than the intrinsic line width, and could be comparable to that of the turbulent flow velocities.  However, thermal broadening can be accommodated by convolving a Gaussian function representing that broadening with the line profiles calculated here.  Equation (27) then becomes
\begin{eqnarray}
  dI(V) = J_0 \delta(V-v_t(t)) dt   = J_0 \delta(f(t)) dt \\
  f(t) \equiv V - v_t(t)
  \end{eqnarray}
  To simplify the expression $\delta(f(t))$, I use the following identity from the algebra of delta functions \citep{Dennery67}
  \begin{equation}
    \delta(f(t)) = \sum_i^N \frac{1}{|\frac{df}{dt}|_{t=t_i}|}\delta(t-t_i)
  \end{equation}
  where $t_i$ is the $i^{th}$ root of $f(t) = 0$.  In the case of the Batchelor Vortex, there will be two roots at $\pm t$ and the same value of $|\frac{df}{dt}|$.
  Substitution of Equation (32) into Equation (30) and integration over $-L/2 \leq t \leq L/2$ yields the expression for the ``observed'' spectrum of the Batchelor Vortex
  \begin{eqnarray}
    I(V) = 2J_0 \left(\frac{1}{|-\frac{dv_t}{dt}|_{t=t_0}|} \right) \mbox{ if } v_t = V \mbox{ at any point where } t=t_0  \\
    = 0 \mbox{ if } v_t \neq V \mbox{ at any point on LOS}
  \end{eqnarray}
  Equation (33) incorporates the fact that, for the Batchelor Vortex, there will be two points along the LOS at $t = \pm t_0$, if there is any point at which $f(t) = 0, V = v_t, \mbox{ for } -L/2 \leq t \leq L/2$.

  For the case of the flow field in the Batchelor Vortex, Equation (26), we have
  \begin{equation}
    \frac{dv_t}{dt} = V_0 \left[ -\frac{2st}{(s^2 + t^2)^2} \left(1 - \exp(-(s^2 + t^2)e^{\tau}) \right) + \frac{s}{s^2 + t^2} \left(2t e^{\tau} \exp(-(s^2 + t^2)e^{\tau}) \right) \right]
  \end{equation}
  The dimensional constant $V_0$ is an overall scale factor that sets the velocity scale of the vortex.  The function Equation (35), substituted into Equation (33), gives the expression for the spectrum.  
  
  With the above equations, the emission spectrum of a Batchelor Vortex can be calculated.  This spectrum depends on the velocity scale of the vortex, $V_0$, the dimensionless age $\tau$, and the scale size of the vortex $a_0$.  In what follows, I take $\tau = 0$; converting to times $\tau \neq 0$ presents no difficulties.  I set $V_0 = 1$, meaning any vortical flow can be studied by multiplying by the appropriate velocity scale.  Finally, I leave the distances $s$ and $t$ dimensionless so they, also, can be scaled to any flow of interest.  
  
  The spectrum of a vortex is calculated using Equations (26), (33), and (35).  The specific steps are as follows. 
  \begin{enumerate}
    \item A velocity model is chosen, using Equation (26), with parameters $V_0 = 1$ and $\tau = 0$ for the results presented here.
  \item The range of Doppler velocities of interest are chosen, ranging from the maximum value $v_t(\mbox{max})$  to  $v_t(\mbox{min})$ (defined in Section 3.3 following Equation (24)).
  \item A set of values of $V$ are chosen $v_t(\mbox{min}) \leq V \leq v_t(\mbox{max})$. These correspond to the ``channels'' in an observed spectrum.  They are chosen to be at equal spacing in $V$ between the limits.  In a typical calculation carried out, the number of channels $N$ was arbitrary chosen to be 31 - 32.  Any value of $N$ could be chosen.
  \item The derivative of the LOS component of the velocity, $\frac{dv_t}{dt}$ is calculated using Equation (35).
  \item For each of the $N$ ``channel velocities'' $V_i, i = 1, N$, the corresponding value of $t_i$ is obtained by solving the equation
    \begin{equation}
      v_t(t_i(V_i)) - V_i = 0
    \end{equation}
    for $t_i$.
    The resulting solution $t_i$ gives the position along the LOS at which $v_t = V_i$, the position at which Doppler-shifted radiation would be emitted at the observed velocity $V_i$
  \item Using this set of $t_i$ values, the value of $|-\frac{dv_t}{dt}|_{t=t_i(V_i)}|$ for each of the channels is calculated, and used in Equations (32) and (33) to calculate the ``observed'' spectrum.  This spectrum is the final output of the analysis.

   An illustration of this calculation is shown in Figure 4. Spectra are shown for two lines of sight with $s = 0.5, 1.0$, i.e. through the central part of the vortex.  
    
  \end{enumerate} 
 \begin{figure}[h!]
\begin{center}
\includegraphics[scale=0.85,angle=0]{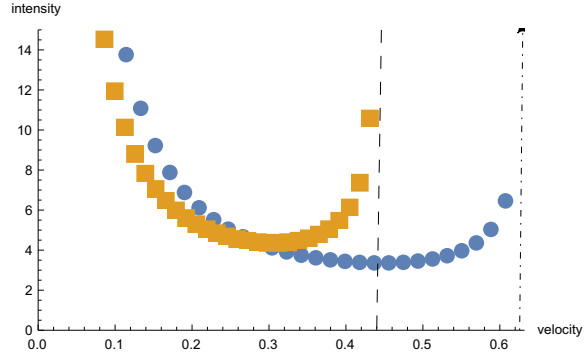}
\caption{The spectrum as given by Equation (33), in units of $2J_0$, for the case of $\tau = 0$ and $V_0 = 1$. The spectrum in gold squares represents the spectrum observed at a dimensionless offset distance from the vortex axis $s = 0.5$.  The spectrum with blue solid circles is the spectrum for  $s = 1.0$. Both lines of sight pass within the vortex, and thus regions of high vorticity.  The dashed vertical line indicates the maximum Doppler shift velocity for the LOS with $s = 0.5$, and the dot-dash vertical line is the limiting velocity for the case $s = 1.0$.  Both spectra show a peak at zero Doppler shift, corresponding to gas outside the vortex, or vortical flow perpendicular to the LOS, a minimum in the emission at intermediate velocities, and a peak at the limiting velocity corresponding to the peak LOS component of the vortical flow. The detailed characteristics of the spectrum, most importantly the peak at the limiting velocity, change with the offset distance $s$. }
\end{center}
\end{figure}

The spectra in both cases show common features; strong emission at low Doppler velocities corresponding to emission from gas that is stationary or slowly moving with respect to the observer, weaker emission at intermediate velocities corresponding to more pronounced motions associated with the vortex, and finally a rise to a ``spectral peak'' at the maximum Doppler velocity $v_t(\mbox{max})$.  There is no emission for velocities higher than this value.  These features can be understood on the basis of the inverse dependence of the intensity on the derivative of $v_t$ (Equation 33), and the form of this function as shown in Figure 3.

The details of the spectrum, most importantly the Doppler velocity of the peak associated with $v_t(\mbox{max})$, depend on the properties of the vortex and the value of the offset distance $s$.  The spectrum in Figure 4 is for $s \geq 0$, i.e. lines of sight on the side of the vortex axis corresponding to $v_t \geq 0$.  As may be seen in Figure 2, for lines of sight at negative $s$, corresponding to the other side of the vortex, the spectrum would be antisymmetric, the same in form but for Doppler velocities which are negative (blueshifted). This analysis thus suggests a possibly simple, qualitative observational test for vortices in astrophysical flows.  Spectral mapping of the vicinity of a vortex might yield a characteristic pattern, qualitatively similar to spectral observations of spiral galaxies (see Section 5 below).

For lines of sight which pass well outside the core of the vortex, Doppler-shifted gas will appear in the spectrum, but the vorticity along the LOS will be close to zero (Equation 18). Figure 5 shows spectra calculated in the same way, and presented in the same format, as Figure 4.  The two spectra correspond to offset (from the vortex axis) distances of 3 and 4, as described in the figure caption.  It is clear that the same qualitative features are present in these spectra as are discussed above for lines of sight through the vortex. This indicates that a quantitative, rather than purely qualitative analysis is necessary to test for the presence of vorticity in emission spectra.   
 \begin{figure}[h!]
\begin{center}
\includegraphics[scale=0.85,angle=0]{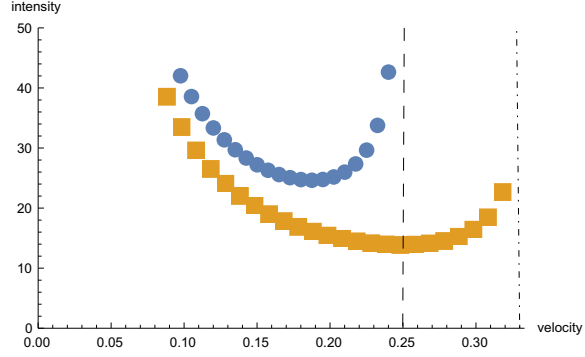}
\caption{ Spectra, in the same format as Figure 4, for line of sight distances of  $s = 3$ (gold squares) and  $s = 4$ (blue circles), outside the vortex where the flow is largely irrotational.  The spectra are qualitatively the same as those measured through the core of the vortex, although the quantitative details differ.  Vertical lines again indicate the limiting velocities for each spectrum. }
\end{center}
\end{figure}
\subsection{3.5 Determination of the ``Pseudovorticity'' from the Spectrum}
To this point, I have discussed the emission spectrum from a vortex of specified properties. and noted the characteristics of that spectrum.  In this section, I consider the astronomically-relevant question of taking a spectrum, which presumably roughly resembles the idealized spectra in Figures 4 and 5, and inferring the vorticity.

The prime observational feature in such a spectrum is the ``spectral line'' at the maximum vortex velocity $v_t(\mbox{max}) = v_t(s,0)$.  As indicated in this expression, this peak is a function of the (dimensionless) coordinate $s$.

The plane containing the line of sight is, by construction, one with Cartesian coordinates $(x,y) \rightarrow (s,t)$.  Retaining the $(x,y)$ coordinates, the $z$ component of the vorticity, which is the only one which is nonzero, is
\begin{equation}
  \omega_z =  \left( \frac{\partial v_y}{\partial x} - \frac{\partial v_x}{\partial y} \right)
\end{equation}
A potentially observable change of $v_t(\mbox{max})$ with $s$ corresponds to the first term on the left side of Equation (37), albeit path-integrated.  However, the second term is completely unknowable, since neither the numerator nor the denominator can be measured.

\cite{Raymond20a,Raymond20b} used a statistical argument that the variances of the (spatially varying) terms on the RHS of Equation (37) would be equal.  They therefore multiplied their measured gradient by $\sqrt{2}$ to obtain their estimate of the vorticity.  In the present study, the vortex responsible for the model spectrum is a deterministic flow, not a random field of vorticity, so a statistical argument is not used.  Instead, it is  noted that the magnitude of the vorticity in the flow field given by Equation (23) depends only on $\sqrt{s^2 + t^2}$, so the two terms in Equation (37) have equal magnitudes.  I there use a symmetry argument to the effect that on average,
\begin{equation}
  |\frac{\partial v_y}{\partial x}| \simeq |\frac{\partial v_x}{\partial y}|
\end{equation}
and that the signs of the two terms would be opposite. I therefore multiply the observed gradient by a factor of 2.  

With the above uncouth assumptions, I introduce an observational estimate of the vorticity in an astronomical fluid, the {\em pseudovorticity} $\omega_{\psi}$, defined as 
\begin{equation}
  \omega_{\psi} \equiv 2 \left( \frac{\Delta v_t(\mbox{max},d)}{\Delta s_d} \right) = \frac{2 V_0}{a_0} \left(  \frac{\Delta v_t(\mbox{max})}{\Delta s}  \right) 
\end{equation}
where the subscript ``d'' in the first terms on the left hand side indicate the dimensional forms of the LOS velocity and transverse coordinate s, and the second set of terms on the left hand side are the dimensionless versions shown in Figures 4 and 5.

I now apply these notions to the spectra shown in Figures 4 and 5.  Begin with Figure 4 for the Batchelor Vortex, applying Equation (39) with values ``measured'' from Figure 4.  For this calculation, $V_0 = a_0 =1$. The change in dimensionless coordinate $\Delta s = 0.5$, and the shift in the dimensionless maximum velocity $\Delta v_t(\mbox{max}) = 0.632 - 0.442 = 0.190$, so $ \omega_{\psi} = 0.76$.

To see if this rude estimate is approximately valid, I compare it to the ``true'' vorticity at dimensionless distances of 0.5 and 1.0 from the vortex axis.  Using Equation (19) for $\omega(s=0.5, t = 0, \tau=0) \mbox{ and } \omega(s=1.0, t = 0, \tau=0)$, I obtain values of 1.55 and 0.74, respectively.  The conclusion for this case is that the observationally-inferred pseudovorticity yields the true value within a factor of 2.

This exercise can be repeated for the ``observations'' of Figure 5, which simulate a pair of lines of sight far from the vortex axis, where the vorticity is close to zero.  Using Equation (19) for the case shown in Figure 5 yields values of  $2.4 \times 10^{-4} \mbox{ and } 2.2 \times 10^{-7}$ for $s = 3,4$ respectively. Application of Equation (39) yields  $\omega_{\psi} = 0.16$.

The calculated pseudovorticity is substantially higher than the true, nearly zero value.  However, it is substantially smaller than that which would be inferred for observed lines of sight through the core of the vortex.  This suggests (to a self-deluded individual) that we could use this type of analysis to distinguish between true high and low vorticity media, and perhaps have a semi-quantitative measure of the vorticity in a fluid.

However, it is obvious that the low vorticity case discussed here concerns an LOS at great distances from the vortex axis, where the azimuthal flow velocity is also substantially lower.  A smaller pseudovorticity would then automatically follow.  What is necessary to check this analysis is modeling an irrotational flow with velocities comparable to those characteristic of the case displayed in Figure 4.  That is the subject of the next section.  
\section{4. The Spectrum and Pseudovorticity of an Irrotational Flow}
To test the reliability of the pseudovorticity as a measurable, it is desirable to consider a flow for which the overall flow scale factor $V_0$ is adjustable, but which is irrotational by construction.  This will be introduced in the following subsection.  I concede at the outset that this might be an academic excercise, in that real astrophysical flows may not conform with mathematical idealizations.   
\subsection{4.1 A Potential Flow}
In this section, I consider a flow which possesses some of the features of the strong vortex of the Batchelor Flow, but which nonetheless has zero vorticity throughout the flow field.  The way to guarantee to this is to have the flow field defined by a potential function $\Phi$, resulting in potential flow,
\begin{equation}
  \vec{v} = \nabla \Phi 
\end{equation}
A flow of this form guarantees $ \nabla \times \vec{v} = 0$.

If we further assume that the flow is incompressible, so $\nabla \cdot \vec{v} = 0$, then the potential function $\Phi(\vec{r})$ satisfies Laplace's equation,
\begin{equation}
  \nabla^2 \Phi = 0
\end{equation}
In selecting the potential function $\Phi$, I pick the same geometry as in Section 3, and initially choose cylindrical coordinates.

This is a standard problem in undergraduate Electricity and Magnetism \citep[e.g.][]{Reitz67}, where it is used to discuss the electric potential and electric field around a conducting cylinder immersed in a uniform electric field.  In the present case, the problem of interest is that of an impermeable cylinder immersed in a uniform flow.  A visualization of the geometry and flow is shown in Figure 6. 
 \begin{figure}[h!]
\begin{center}
\includegraphics[scale=0.85,angle=0]{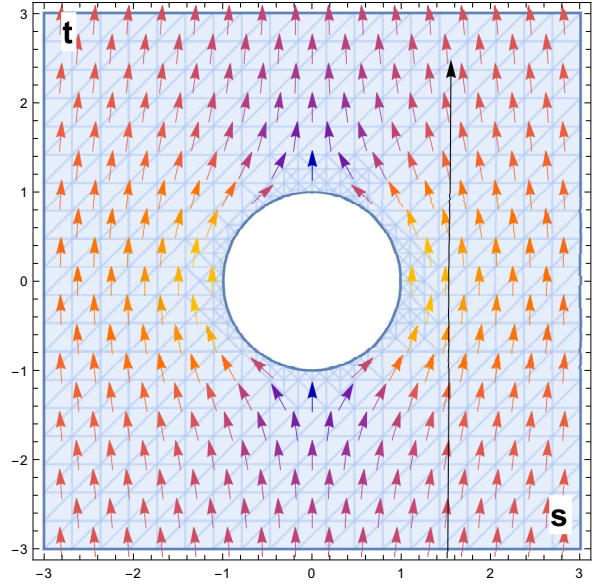}
\caption{Figure, similar in format to Figure 2, but showing the flow field corresponding to a potential flow.  The vertical ray corresponds to an LOS with $s=1.5$.  The hue of the plotted vectors conveys the magnitude of $\vec{v}$. The scale of the system is determined by the radius of the effective cylinder, $R_0$, which is taken to be 1 in this figure.}
\end{center}
\end{figure}

The solution to Laplace's Equation in cylindrical coordinates can be expressed in a series expansion of terms characterized by a ``quantum number'' $n$ \citep{Reitz67}
\begin{equation}
  \Phi(R,\phi) = A_0 + B_0 \ln R + A_1^c R \cos \phi + A_1^s R \sin \phi + B_1^c \left( \frac{1}{R} \right) \cos \phi + B_1^s \left( \frac{1}{R} \right) \sin \phi + \ldots
\end{equation}

The solution to the problem of interest, i.e. the appropriate set of coefficients in Equation (42), is found by incorporating the boundary conditions.  In this case, boundary conditions are
\begin{eqnarray}
  \vec{v}(\vec{r} \rightarrow \infty) = +V_0 \hat{e}_t \\
  v_n (R = R_0) =  v_R (R = R_0) = 0
\end{eqnarray}
Where $\hat{e}_t$ is the unit vector in the Cartesian $t$ coordinate introduced in Section 3.3, and $v_n (R = R_0)$ is the component of the velocity normal to the surface of the cylinder at the surface of the cylinder. It should be noted that the inner boundary condition, Equation (44), differs from that used in undergraduate studies of electrostatics.  It is also worth emphasizing that the ``cylinder'' is a mathematical artiface to produce a flow with roughly realistic properties, and need not correspond to an actual astronomical object.

To make an irrotational flow field which simulates that of a vortical flow with similar geometry to that considered in Section 3, I define $\phi = 0$ to be along the $t$ axis in the direction of $t \rightarrow - \infty$, and flowing in a counterclockwise direction for $s > 0$ (see Figure 6).

The boundary conditions (43) and (44) are satisfied if the coefficients in Equation (42) satisfy the following:
\begin{equation}
  A_1^s = B_0 =0; A_1^c = -V_0; B_1^s = 0; B_1^c = R_0^2 A_1^c = -V_0 R_0^2
\end{equation}
All higher order coefficients, e.g. $B_n^s, A_n^c$, for $n > 1$, are zero.

With the boundary conditions selected, the solution for the potential function $\Phi$ is
\begin{equation}
  \Phi (R, \phi) = -V_0 \left[ R + \left( \frac{R_0^2}{R} \right)  \right] \cos \phi
\end{equation}
The corresponding velocity field, in cylindrical coordinates, is
\begin{equation}
  \vec{v} (R, \phi) = \hat{e}_R \left[ -V_0 \cos \phi + V_0  \left( \frac{R_0^2}{R^2} \right) \cos \phi \right]
  +  \hat{e}_{\phi} \left[ V_0 \sin \phi + V_0  \left( \frac{R_0^2}{R^2} \right) \sin \phi \right]
  \end{equation}

  To compare this flow with that of the Batchelor Flow, I write the flow velocity in the Cartesian coordinate system of Figure 2,
  \begin{equation}
    \vec{v} (s,t) = v_s(s,t) \hat{e}_s + v_t(s,t) \hat{e}_t
  \end{equation}
  where $s,t$ are again dimensionless coordinates, de-dimensionalized like Equation (25), except in the present case the dimensional coefficient is $R_0$, the radius of the effective cylinder rather than the Gaussian radius of the Batchelor Vortex.

  The resulting components of the flow velocity are
  \begin{eqnarray}
    v_s(s,t) = - \frac{2V_0 st}{(s^2 + t^2)^2}\\
     v_t(s,t) = V_0 \left[1 +  \frac{s^2 - t^2}{(s^2 + t^2)^2} \right]
  \end{eqnarray}

  This flow field, which will be used in our analysis of the model spectrum and deduction of the pseudovorticity for an irrotational flow field, is plotted in Figure 6. The flow speed is higher than $V_0$ close to the cylinder, a consequence of the Equation of Continuity.

  The flow field in Figure 6 shows a degree of similarity with the Batchelor Flow; for $s > 0$ there is an azimuthal flow in the counterclockwise direction.  In the next subsection, I examine the degree to which the spectrum emergent from this flow resembles that from the Batchelor Vortex discussed in Section 3.  
  \subsection{4.2 Potential Flow: Spectrum and Pseudovorticity}
  To calculate the spectrum of radiation, and deduce the pseudovorticity for the case of the potential flow field, I follow the same steps as in Section 3 for the Batchelor Flow.  Equation (32) is general, and is used here.  That equation shows that the observed spectrum is determined by the profile of $v_t(s,t)$ for fixed s.  The spectrum is dominated by ``emission lines'' at velocities which correspond to zeros of $\frac{dv_t}{dt}$.

The expression for $v_t(s,t)$ used here is Equation (50), and it is plotted as a function of $t$ in Figure 7.
\begin{figure}[h!]
\begin{center}
\includegraphics[scale=0.85,angle=0]{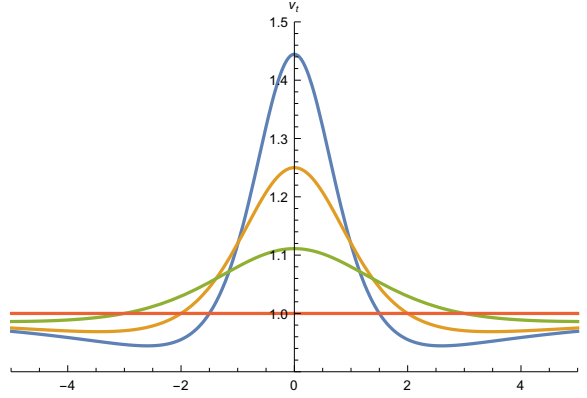}
\caption{The line-of-sight component of the fluid velocity ($v_t$) for the case of potential flow given by Equations (49) and (50) as a function of $t$, the coordinate along the line of sight.  For all three curves, the normalization constant of the velocity $V_0 = 1.0$, and the dimensionless thickness of the emitting region $L = 10$. The three curves give $v_t$ for spatial offset values (distance from the center of the effective ``cylinder'', dimensionless) of $s=1.5$ (blue), $s=2.0$ (yellow), and $s=3.00$ (green). As in the case of the Batchelor Vortex shown in Figure 3, and for all three offset distances, it is seen that the the line-of-sight component of the velocity maximizes at $t=0$, and diminishes to a low value near the boundary of the emitting region. However, in contrast to the Batchelor Vortex, the LOS component $v_t$ has another extremum at an intermediate value of $t$. }
\end{center}
\end{figure}
Figure 7 shows that for $s > 0$, there are three values of t for which $\frac{dv_t}{dt} = 0$, as opposed to two for the Batchelor Vortex (Figure 3).  These values are as follows.
\begin{enumerate}
\item For $|t| \gg 1, \frac{dv_t}{dt} \rightarrow 0$.  This corresponds to the ``unperturbed flow''.
\item For intermediate values $2 \leq t \leq 4$, $v_t < V_0$.  This corresponds to ``steering'' or ``deflection'' of the flow away from the line of sight. In this intermediate range, there is a minimum value of $v_t$ (``blueshifted'' with respect to $V_0$) and a corresponding case of $\frac{dv_t}{dt} = 0$.  As shown in Figure 7, the location of this minimum value of $v_t$ depends on $s$.  
\item Finally, for $t \rightarrow 0$, $v_t$ increases to  $v_t > V_0$, reaching a maximum at $t = 0$, as is the case for the Batchelor Vortex.  This maximum is a consequence of the equation of continuity as the flow is blocked by the cylinder, although I again emphasize that the literal model of the flow field is not the point of interest here.
\end{enumerate}

  Figure 7 allows us to sketch a qualitative spectrum of the emission from the potential flow, as shown in Figure 8.  
\begin{figure}[h!]
\begin{center}
\includegraphics[scale=0.85,angle=0]{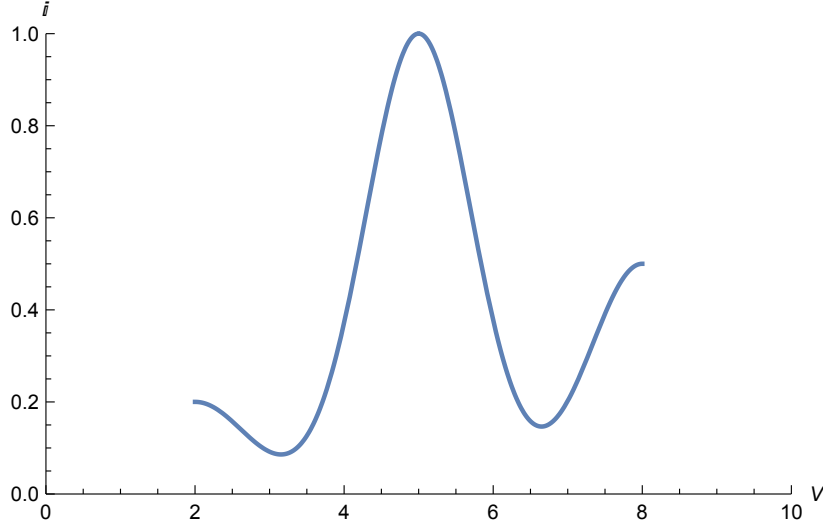}
\caption{Heuristic representation of the spectrum for the potential flow displayed in Figure 6.  The spectrum from such a flow would show three peaks: a peak corresponding to flow unshifted with respect to the net flow at $V_0$ (central peak), a blueshifted peak when the flow is being diverted away from the line of sight (smaller peak to the left), and a strong redshifted peak when the LOS speed is higher than $V_0$ around $t=0$ (larger peak on right). The form of this spectrum could be inferred from Figures 6 and 7 and Burton's insights on Galactic structure from 21 cm spectra (see Section 5).   }
\end{center}
\end{figure}
The large central peak (corresponding to $V = V_0 =5$) is undeflected and unperturbed gas.  The dominance of this feature is determined by the size scale of the total emitting region ($L$) relative to the size scale of the flow perturbation  ($R_0$).  The peak at lower velocities ($V \simeq 2$) corresponds to flow that is ``deflected but not accelerated''.  The absence of emission at velocities lower than this peak ($V < V_{min}$) is complete. Finally, the peak with $V = V_{max} > V_0$ is dominated by emission at $t \simeq 0$.  Again, there is no emission at $V > V_{max}$.

As indicated in Equations (33) and (34), the spectrum is determined by the function $\frac{dv_t}{dt}$.  For the potential flow field (47) - (50), this function is
\begin{equation}
 \frac{dv_t}{dt} = \left( \frac{V_0}{R_0} \right) \frac{-2t}{(s^2 + t^2)^2} \left[ 1 + \frac{2 (s^2 - t^2)}{(s^2 + t^2)} \right] 
\end{equation}
The leading term on the right hand side makes the overall expression dimensionally correct.  The variables $s,t$ in Equation (51) are dimensionless, as in Section 3.

In what follows, I carry out the analysis for the ``redshifted'' feature of the generic spectrum of Figure 8, corresponding to the flow around $t \sim 0$. This arises from a part of the flow which crudely resembles the Batchelor Vortex in the sense of consisting of an azimuthal flow.  Emission of this feature will be present in the velocity range $V_0 \leq V \leq V_{max}$.  As in the case of the Batchelor Vortex, two values of $t$ contribute to each value of $V$ in this range.

Spectra for the potential flow field, calculated with Equations (33) and (34) are shown in Figure 9, for two values of s.  
 \begin{figure}[h!]
\begin{center}
\includegraphics[scale=0.85,angle=0]{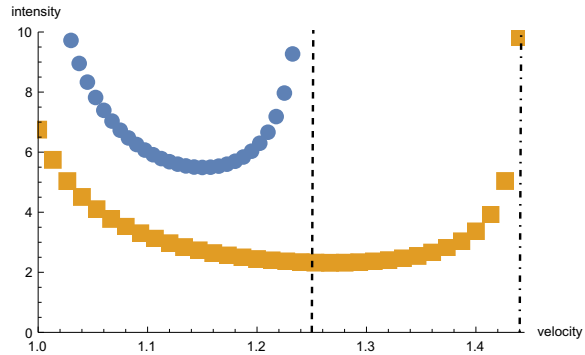}
\caption{ Spectra, in the same format as Figures 4 and 5, for the case of the potential flow discussed in Section 4.1. Different colored symbols correspond to line of sight distances of  $s = 1.5$ (gold squares) and  $s = 2$ (blue filled circles). Vertical lines again indicate the limiting velocities for each spectrum.  The spectra show the same qualitative features as those of the Batchelor Vortex (Figure 4), although the vorticity is identically zero in the case shown here.}
\end{center}
\end{figure}
What is immediately obvious from Figure 9 is that the spectra have the same qualitative form as Figure 4 for the Batchelor Vortex.  This similarity is a consequence of Eq (33,34) and that fact that, for a ``circulating flow'', there is a point in the flow for which $\frac{dv_t}{dt} = 0$.  The principal result of Figure 9 is that this flow field, like the Batchelor Vortex, produces a spectrum with a peak at low velocities ($V = V_0$), a fall-off to a minimum intensity at intermediate velocities, then a rise to a ``spectral line'' at $V = V_{max}$.

I now proceed to calculate the pseudovorticity in the same way as in Section 3, utilizing Equation (39).  Once again, for the spectra in Figure 9, the dimensional factors $V_0, R_0$ are taken to be unity.  For the spectra in Figure 9, $\Delta v_t(max) = 0.19$ and $\Delta s = 0.50$, yielding a pseudovorticity $\omega_{\psi} = 0.76$.  This is {\em exactly} the pseudovorticity inferred from the spectrum produced by the Batchelor Vortex, even though in the former case the flow field consisted of an idealized vortex, and in the present case the flow field is mathematically constructed to possess zero vorticity.

An extreme statement of this result is that a path-integrated spectrum, providing information on only one component of a three-dimensional flow field, does not preserve or convey information on the vorticity in the fluid.  Insight into this result is given in the next section. Before leaving this section, it should be pointed out that spectroscopic mapping observations could, in principle, distinguish between the idealized flows considered here.  As noted in Section 3.4, in the case of the Batchelor Vortex (or any other simple vortical flow), the Doppler shifts for $s < 0$ would be equal in magnitude and opposite in sign to those for  $s > 0$.  In the case of the potential flow considered immediately above, spectra at $\pm s$ would be the same (Doppler shifts equal in magnitude and sign).   
\section{5. Vorticity Measurements from Spectroscopy and Burton's Admonitions on Galactic Structure from Spectroscopy}
\begin{center}
{\em ``There is great unity in physics''} \footnote{The first sentence in the document for a laboratory exercise I did while an undergraduate.  I forget what the experiment was, but remember the inspiring invocation.}
\end{center}
The conclusion of the paper to this point is that spatially-resolved  spectra cannot yield a certain measurement of the vorticity in a fluid.  While a test case in which vorticity is present, the Batchelor Vortex, yields satisfactory results, a velocity field constructed to have zero vorticity appears to have similar vorticity.  I have claimed that any ``circulating'' flow, that is one with a substantial azimuthal flow over part of the fluid, would yield the observational signature of vorticity shown in Figures 4, 5, and 9.  

This result could actually have been anticipated from work in a very different area of astronomy, the study of the distribution of hydrogen gas in the Galaxy from observations of the 21 cm line. This similarity motivated the statement about the unity of physics given at the start of this section.  In the early and mid 1970s, Butler Burton of the National Radio Astronomy Observatory pointed out that integrated (through a long line of sight through the Galaxy) 21 cm spectra could give the impression of spatially-localized regions of hydrogen, even when the hydrogen gas was uniformly distributed.    Burton pointed out that, even if the gas were uniformly distributed along the line of sight rather than clumped in clouds or spiral arms, bright emission at a specific velocity would be observed if there were small changes in the line-of-sight velocity for large distances along the LOS \citep{Burton71}.  Fainter emission would be measured at other velocities. This situation could arise in a number of ways, but one of the more obvious was for the gas interior to the orbit of the Sun being in circular motion.  Burton pointed out that the gradient in LOS velocity with distance along the LOS was small in the vicinity of the tangent point.  \cite{Burton71} succinctly stated the effect as {\em ``\ldots for any reasonable rotation law, the velocity observed from regions near the subcentral point changes very slowly along the line of sight, so that the profile contains a contribution from a long pathlength near the maximum velocity.  This effect results in the high-velocity ridge pattern characteristic of both the observed and the model contour maps.''}.  Another comment in this paper is that {\em `` \ldots intensities at velocities at which $|\frac{dV}{dr}|$} (the equivalent of $\frac{dv_t}{dt}$ in the present paper) {\em is small will be enhanced on the profiles''}.  

This effect is also mentioned in the magisterial review of Galactic HI observations and interpretation by Dickey and Lockman \citep{Dickey90}, where the term ``velocity crowding'' is introduced, defined as ``\ldots a long distance can be ``crowded'' into a small velocity interval''.  The net effect is that emission from a large distance is concentrated in a small range of observed velocities, resulting in a prominent ``spectral line'' in the spectrum.

These results from studies of the Galactic-scale distribution and kinematics of Galactic hydrogen are highly analogous to the case I have discussed for turbulent, vortical motion in astrophysical fluids that are emitting  optically-thin line emission.  Those who wish to carry out turbulence measurements from spectroscopy must be aware of the admonition in \cite{Burton71}, ``any interpretation of the observations must deal with these geometrical effects ''.  
\section{6. Summary and Conclusions}   
\begin{enumerate}
\item A vortex viewed perpendicular to the axis of vorticity will produce an emission spectrum, in the optically-thin case, which would allow the vorticity in the fluid to be retrieved from astronomical measurements.  This statement is predicated on the line of sight containing a single vortex which is resolved by the astronomical observations, say the beam of a radio telescope. 
\item In this optimum case, the estimate of vorticity which can be retrieved from the observations, with assumptions about the symmetry of the flow, is a reasonable estimate of the true vorticity. 
\item Fluid flows which are {\em irrotational}, i.e. having zero vorticity, but possessing azimuthal flow, or at least a radius of curvature perpendicular to the streamlines for some of the distance along the line of sight, can also yield an observed spectral signature that would indicate  the presence of vorticity in the radiating medium.  In this case, the observations would yield a finite value whereas the true vorticity is zero.  
\item As a result, remote sensing detections of fluid vorticity must be considered ambiguous.
\item It is beyond the scope of this paper to consider the obvious fact that real observations of a turbulent fluid will generally include contributions of numerous vortices of different magnitude, sign of vorticity, and orientation with respect to the line of sight.  One suspects that the aggregate spectrum would contain even less information on the properties of the vorticity field than the idealized case of a single vortex considered here.  
\item The idealized model spectra calculated here (Figures 4, 5, and 9) would suggest that observational searches for vorticity analyse the entire line profile (if accessible) rather than rely on measurement of line centroids.  The full spectrum might permit identification of features corresponding to $\frac{dv_t}{dt} \simeq 0$.
\end{enumerate}
\begin{acknowledgements}
  I thank John Raymond of the Harvard-Smithsonian Center for Astrophysics for reading the first version of this paper, and providing interesting and helpful comments.  
  \end{acknowledgements}

\end{document}